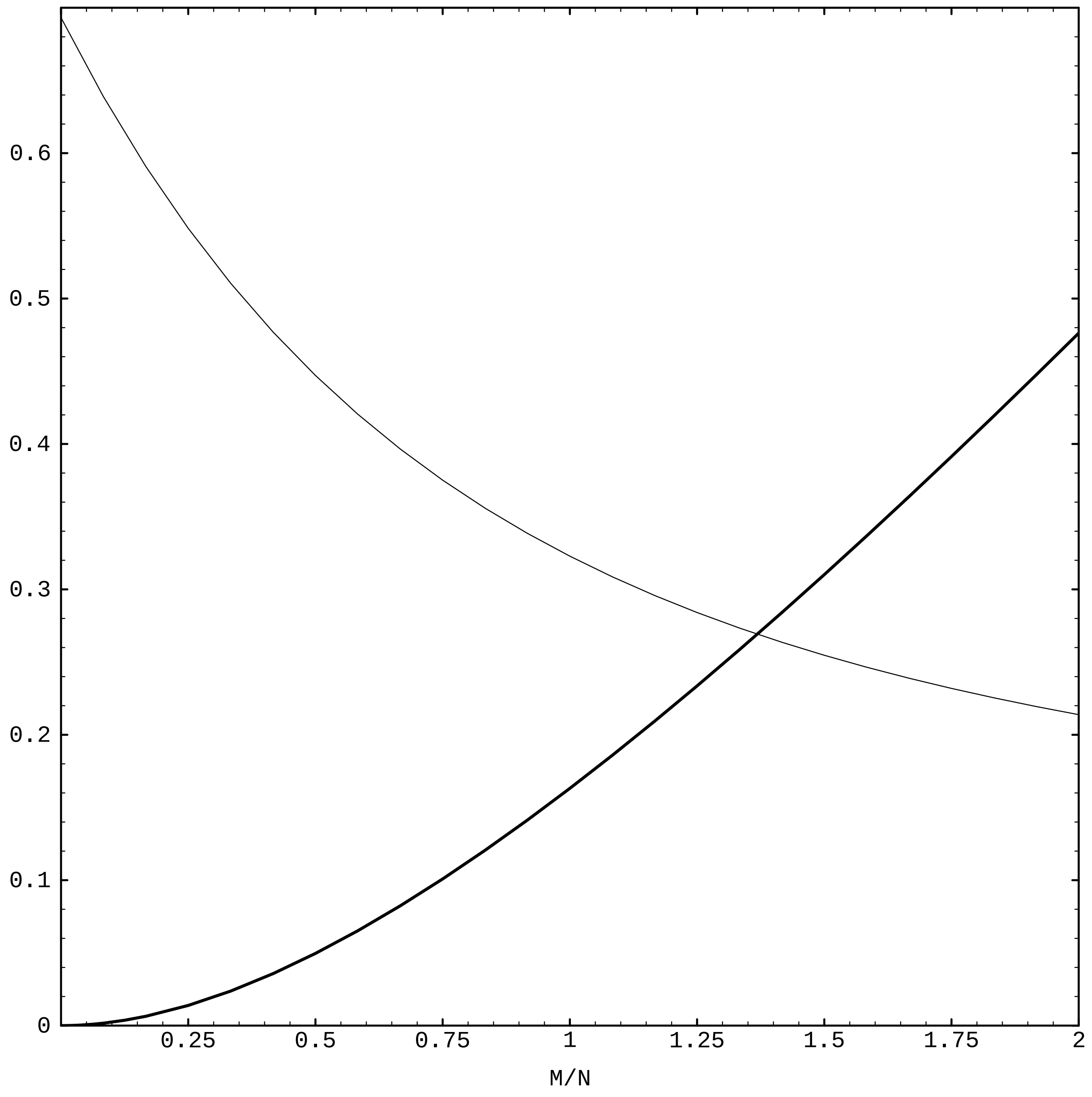

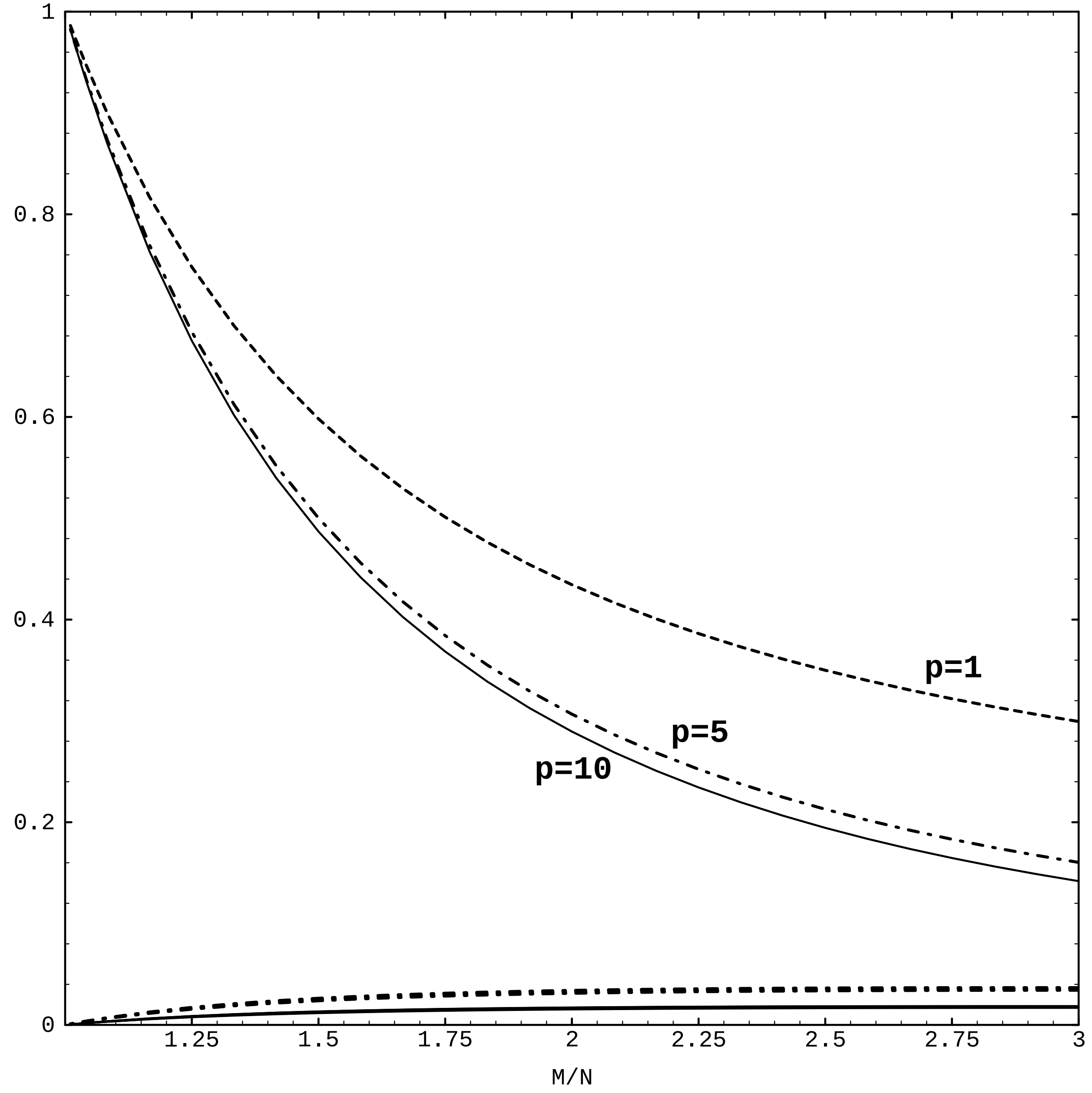

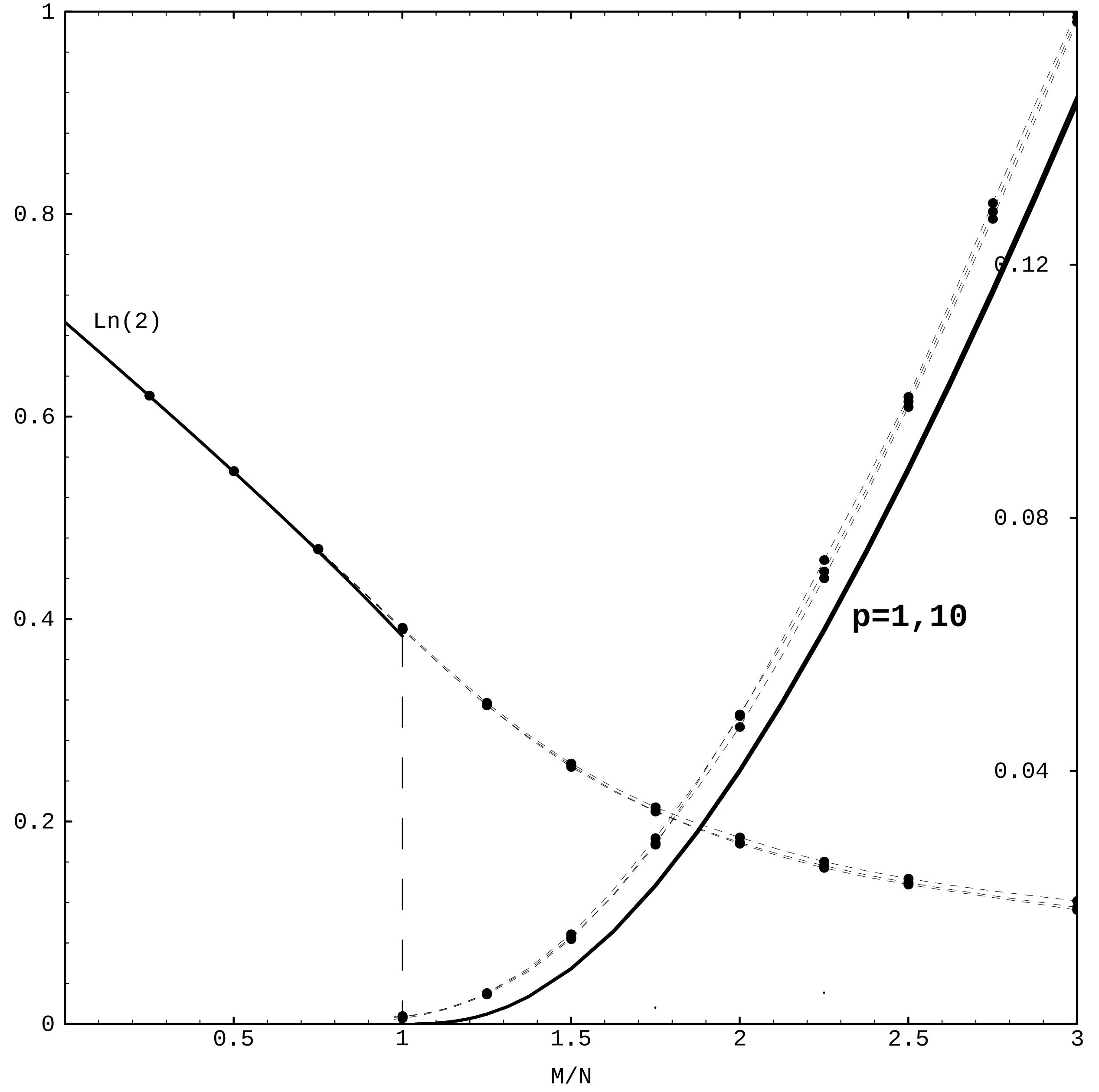

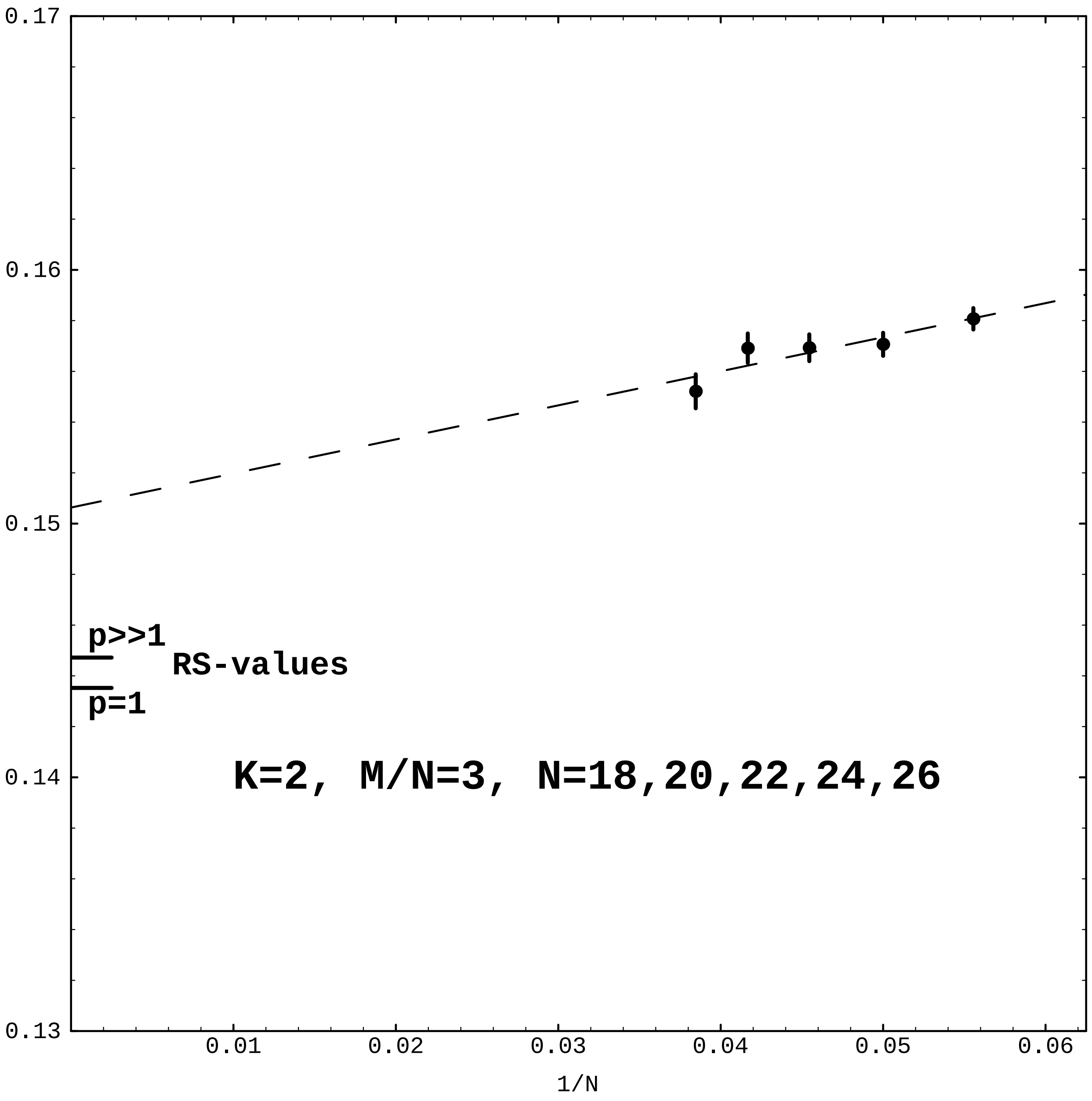

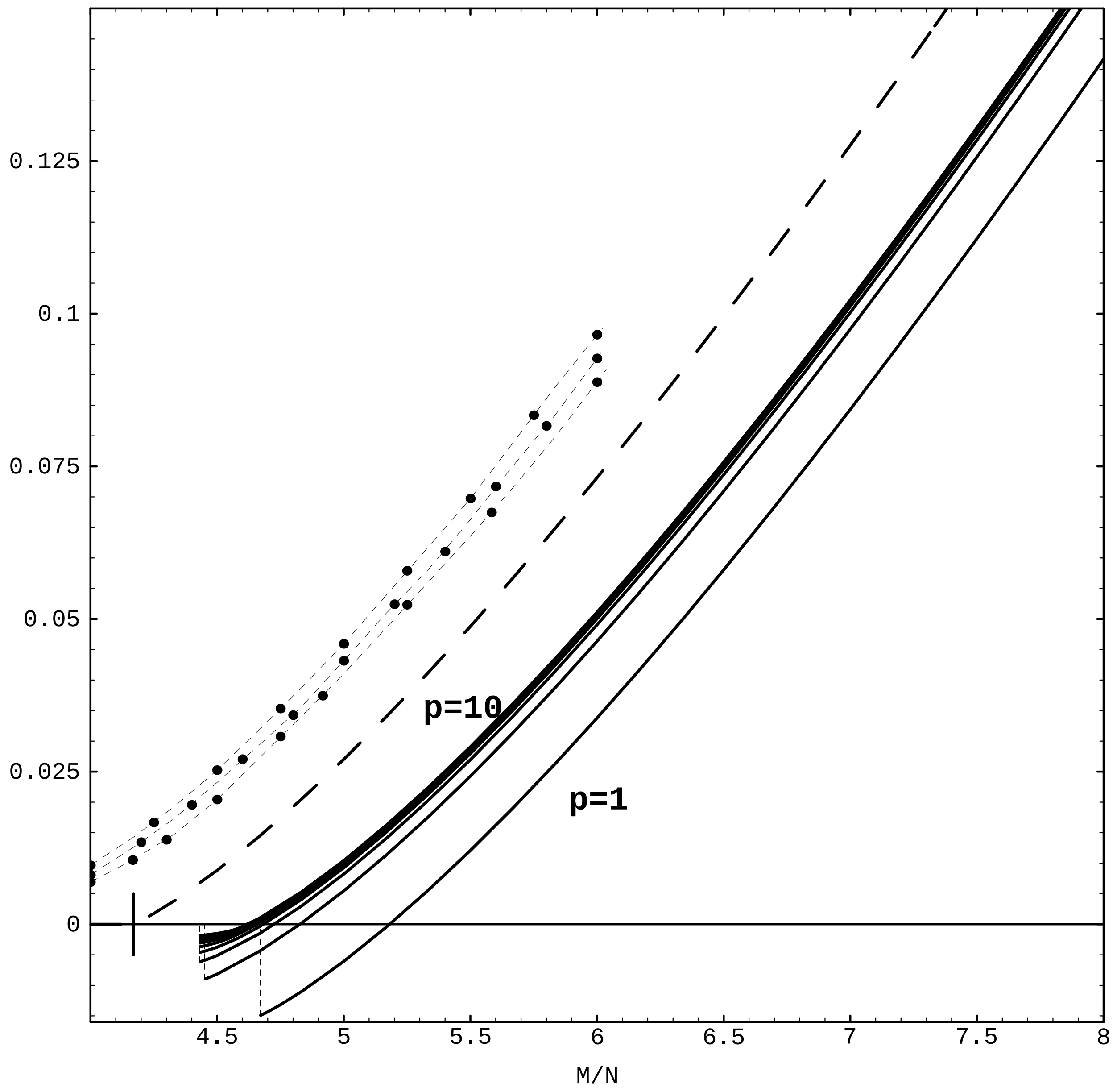

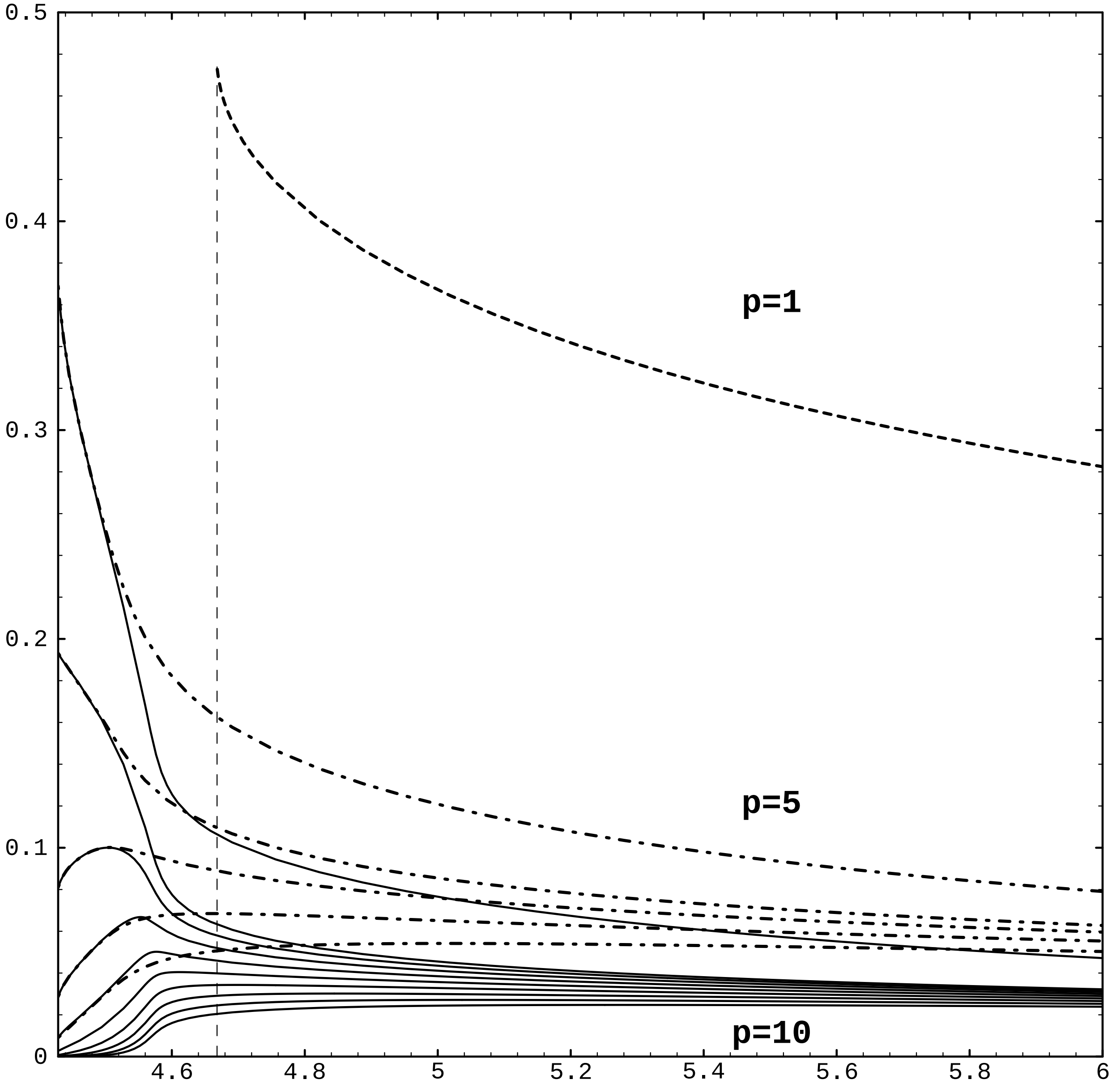

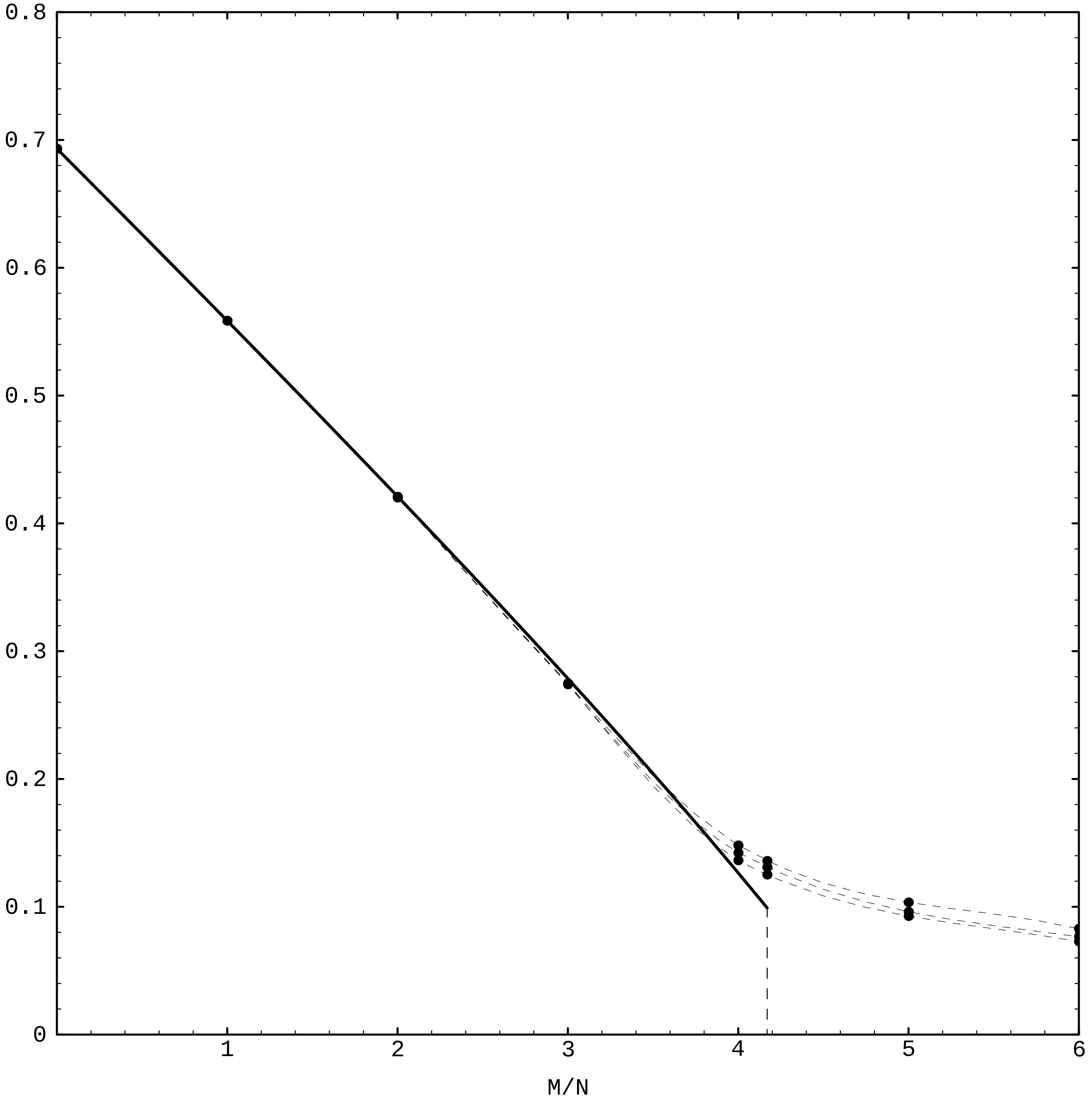

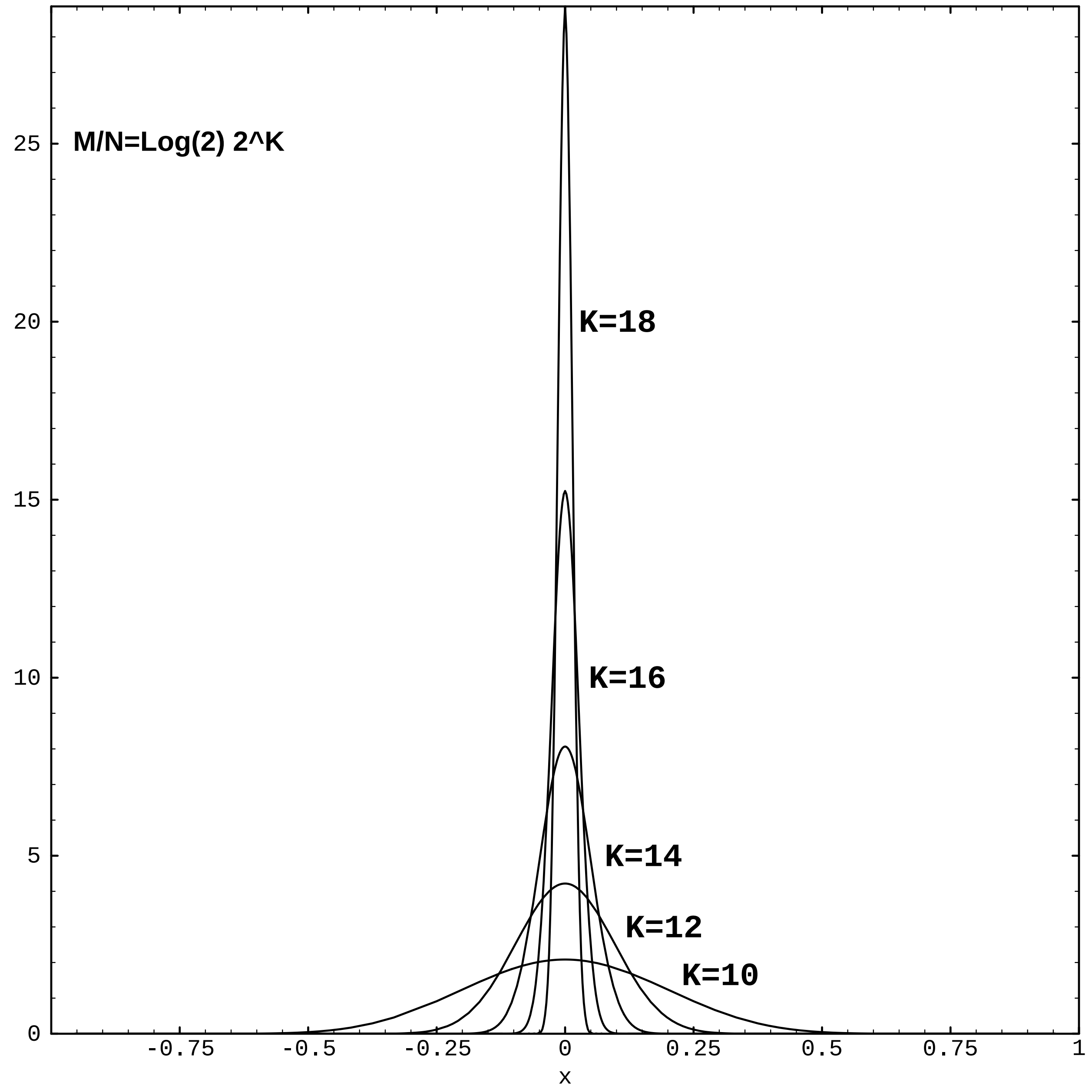

# Statistical Mechanics of the Random $K$-SAT Model

Rémi Monasson* and Riccardo Zecchina†

* *Laboratoire de Physique Théorique de l'ENS,*

*24 rue Lhomond, 75231 Paris cedex 05, France*

† *Dipartimento di Fisica, Politecnico di Torino,*

*C.so Duca degli Abruzzi 24, I-10129 Torino, Italy*

## Abstract

The Random K-Satisfiability Problem, consisting in verifying the existence of an assignment of $N$ Boolean variables that satisfy a set of $M = \alpha N$ random logical clauses containing $K$ variables each, is studied using the replica symmetric framework of diluted disordered systems. We present an exact iterative scheme for the replica symmetric functional order parameter together for the different cases of interest $K = 2$, $K \geq 3$ and $K \gg 1$. The calculation of the number of solutions, which allowed us [Phys. Rev. Lett. 76, 3881 (1996)] to predict a first order jump at the threshold where the Boolean expressions become unsatisfiable with probability one, is thoroughly displayed. In the case $K = 2$, the (rigorously known) critical value ($\alpha = 1$) of the number of clauses per Boolean variable is recovered while for $K \geq 3$ we show that the system exhibits a replica symmetry breaking transition. The annealed approximation is proven to be exact for large $K$.

PACS Numbers : 05.20 - 64.60 - 87.10

# I. INTRODUCTION

The emergent collective behaviours observed in a variety of models of statistical mechanics and in particular in frustrated disordered systems, have been recognized to play a relevant role in apparently distant fields such as theoretical computer science, discrete mathematics and complex systems theory [1–5]. Computationally hard problems, characterized (in worst cases) by exponential running time scaling of their algorithms or memory requirements, the so called NP–complete problems [6], are known to be in one–to–one correspondence with the ground state properties of spin–glass like models (see [1] and references therein). As a consequence, tools and concepts of statistical physics have shed some new light onto the notion of the typical complexity of NP-complete problems and have lead to the definition of new search algorithms as the simulated annealing algorithm, based on the introduction of an artificial temperature and some cooling procedures [7].

Very recently, other techniques inspired from statistical mechanics, namely finite size scaling analysis, have been applied [8] also to the study of universal behaviour in the computational cost (running time) of some classes of algorithms in the course of searching for solutions of random realizations of the prototype of NP–complete problems, the *satisfiability* (SAT) problem we shall discuss.

More generally, phase transition concepts are starting to play a relevant role in theoretical computer science [4], where the analysis of general search methods applied to various classes of hard computational problems, characterized by a large number of relevant variables and generated according to some probability distributions, is of crucial importance in building a theory for the typical–case complexity. NP–complete decision problems which are computationally hard in the worst case appear not to be really so in the typical case, except in critical regions of their parameter space (with a polynomial–exponential pattern) where almost all instances of the problems become computationally hard to solve. Far from criticality, the problems are either under- or over-constrained and both the stochastic search procedures and the systematic ones are capable of finding solutions in polynomial times.

One of the major theoretical open questions in this context would be to understand how typical–case complexity theory of computer science and spin-glass transitions, the so–called replica symmetry breaking transition [1], are related. In turn, computer science is a source of highly non–trivial models containing all the paradigms necessary to a deeper understanding of the physical properties of disordered frustrated systems, in particular diluted models for which the theoretical framework is still to be completed [9,11–15].

Among the known NP–complete problems, the SAT problem is at the same time the root problem of complexity theory [6] and a prototype model for phase transition in random combinatorial structures [3,16]. SAT was the first problem proved to be NP–complete by S. Cook in 1971 [17] and opened the way to the identification of a vast family of other NP–complete problems for which a polynomial reduction to SAT became available [6]. In particular the K–satisfiability (K–SAT) problem, a version of SAT we shall discuss in great detail in what follows, beside playing a central role in NP–completeness proving procedures [6], is a widespread test for the evaluation of performance of combinatorial search algorithms, due the typical intractability of random samples generated near criticality.

In a recent work [5], we have shown that the methods of statistical mechanics of random systems allow to compute some algorithmically relevant quantities such as the typical entropy of the problem, i.e. the typical number of its solutions, and to clarify the nature of the threshold behaviour. The scope of this paper is twofold. On the one hand, we aim at giving a thorough discussion of the analytical derivation of the above results, mainly the calculation of the entropy jump at the transition. On the other hand, we expose in detail the replica symmetric theory of the K–SAT problem by showing both how to go beyond the simplest solution proposed in our previous work [5] and by clarifying the connections with known results in statistical mechanics of diluted models.

The paper is organized as follows. Section **II** is devoted to the presentation of the K–SAT problem and of the known exact results. Sections **III** contains an outline of the statistical mechanics approach whereas the replica symmetric solutions are exposed in Section **IV**. In the successive sections, from **V** to **VIII**, the outcomes of the analytical calculations are exposed in detail for the different values of $K$ of interest. In Section **IX**, we show how to rederive some of the previous results using a simple probabilistic approach. Finally, in Section **X,** some new perspectives opened by the introduction of a model which interpolates smoothly between 2–SAT and 3–SAT are briefly discussed.

## II. THE K–SAT PROBLEM AND A BRIEF SURVEY OF KNOWN RESULTS

Given a set of $N$ Boolean variables $\{x_i = 0, 1\}_{i=1,\ldots,N}$, we first randomly choose $K$ among the $N$ possible indices $i$ and then, for each of them, a literal $z_i$ that is the corresponding $x_i$ or its negation $\bar{x}_i$ with equal probabilities one half. A clause $C$ is the logical OR of the $K$ previously chosen literals, that is $C$ will be true (or satisfied) if and only if at least one literal is true. Next, we repeat this process to obtain $M$ independently chosen clauses $\{C_\ell\}_{\ell=1,\ldots,M}$ and ask for all of them to be true at the same time, i.e. we take the logical AND of the $M$ clauses thus obtaining a Boolean expression in the so called Conjunctive Normal Form (CNF). The resulting K–CNF formula $F$ may be written as

$$F = \bigwedge_{\ell=1}^{M} C_\ell = \bigwedge_{\ell=1}^{M} \left( \bigvee_{i=1}^{K} z_i^{(\ell)} \right) \quad , \tag{1}$$

where $\bigwedge$ and $\bigvee$ stand for the logical AND and OR operations respectively.

A logical assignment of the $\{x_i\}$'s satisfying all clauses, that is evaluating $F$ to true, is called a solution of the K–satisfiability problem. If no such assignment exists, $F$ is said to be unsatisfiable.

When the number of clauses becomes of the same order as the number of variables ($M = \alpha\ N$) and in the large $N$ limit – indeed the case of interest also in the fields of computer science and artificial intelligence [16,18] – the K–SAT problem exhibits a striking threshold phenomena. Numerical experiments have shown that the probability of finding a correct Boolean assignment falls abruptly from one down to zero when $\alpha$ crosses a critical value $\alpha_c(K)$ of the number of clauses per variable. Above $\alpha_c(K)$, all clauses cannot be satisfied any longer and one gets interested in minimizing the number of unsatisfiable clauses, which is the optimization version of K–SAT also referred to as MAX–K–SAT. Moreover, near $\alpha_c(K)$, heuristic search algorithms get stuck in non–optimal solutions and a slow down effect is observed (intractability concentration). On the contrary, far from criticality heuristic processes are typically rather efficient [8].

Very schematically, the known results on K–SAT which have been obtained in the framework of complexity theory may be summarized as follows.

- For $K = 2$, 2–SAT belongs to the class P of polynomial problems [19]. P is defined as the set of computational problems whose best solving algorithms have running times increasing polynomially with the number of relevant variables [6]. For $\alpha > \alpha_c$, MAX–2–SAT is NP–complete [6] : NP–complete problems are the hardest nondeterministic polynomial problems, whose solutions may be found by the exhaustive inspection of a decision tree of logical depth growing in a polynomial way with the number of relevant variables; it is generally thought that the running times of their best solving algorithms scale exponentially with the number of relevant variables [6]. The mapping of 2–SAT on directed graph theory [20] allows to derive rigorously the threshold value $\alpha_c = 1$ and an explicit 2–SAT polynomial algorithm working for $\alpha < \alpha_c$ has been developed [19].

- For $K \geq 3$, both K–SAT and MAX–K–SAT belong to the NP–complete class. Only upper and lower bounds on $\alpha_c(K)$ are known from a rigorous point of view [18,21,22]. Finite size scaling techniques have, recently, allowed to find precise numerical values of $\alpha_c$ for $K = 3, 4, 5, 6$ [3].

- For $K \gg 1$, clauses become decoupled and an asymptotic expression $\alpha_c \simeq 2^K \ln 2$ can be easily found. It is not yet known whether this scaling law is correct or not from a rigorous point of view.

For brevity, we do not discuss here the results concerning the algorithmic approaches to K–SAT and MAX–K–SAT [19,23,24]. We just mention that MAX–K–SAT belongs to the subclass of NP–complete problems which allows for a polynomial approximation scheme for quasi–optimal solutions [23]. A recent numerical study of the critical behaviour in the computational cost of satisfiability testing can be found in [8].

For $\alpha = \frac{M}{N} > 0$, K–SAT can be cast in the framework of statistical mechanics of random diluted systems by the identification of an energy–cost function $E(K, \alpha)$ equal to the number of violated clauses [5,16]. The study of its ground state allows to address the optimization version of the K–SAT problem as well as to characterize the space of solutions by its typical entropy, i.e. the degeneracy of the ground state. The vanishing condition on the ground state energy for a given $K$, corresponds to the existence of a solution to the K–SAT problem and thus identifies a critical value $\alpha_c(K)$ of $\alpha$ below which random formulas are satisfiable with probability one. For $\alpha > \alpha_c(K)$, the ground state energy becomes non zero and gives information on the maximum number of satisfiable clauses, i.e. on the MAX–K–SAT problem. Previous works on the statistical mechanics of combinatorial optimization problems - like traveling salesman, graph partitioning or matching problems [1,25,10,11,9] - focused mainly on the comparison between the typical cost of optimal configurations and the algorithmic results. The issues arising in K–SAT are of different nature, and the key quantity to be discussed [21] is rather the typical number of existing solutions, i.e. the ground state typical entropy $S_K(\alpha)$.

A crucial rigorous result on which the whole statistical mechanics approach is founded concerns the self-averageness taking place in MAX–K–SAT. For any $K$, independently of the particular but randomly chosen sample of $M$ clauses, the minimal fraction of violated clauses is narrowly peaked around its mean value when $N \to \infty$ at fixed $\alpha$ [24].

## III. STATISTICAL MECHANICS OF THE K–SAT AND MAX–K–SAT COST FUNCTION

As discussed above, we map the random SAT problem onto a diluted spin energy–cost function through the introduction of spin variables, $S_i = 1$ if the Boolean variable $x_i$ is true, $S_i = -1$ if $x_i$ is false. The clauses structure is taken into account by a $M \times N$ quenched random matrix $\Delta$ where $\Delta_{\ell,i} = -1$ (respectively $+1$) if clause $C_l$ contains $\bar{x}_i$ (resp. $x_i$), 0 otherwise. Then the function

$$E[\Delta, S] = \sum_{\ell=1}^{M} \delta \left[ \sum_{i=1}^{N} \Delta_{\ell,i} \, S_i ; -K \right] \tag{2}$$

where $\delta[i;j]$ denotes the Kronecker symbol, turns out to be equal to the number of violated clauses in that the quantity $\sum_{i=1}^{N} \Delta_{\ell,i} \, S_i$ equals $-K$ if and only if all Boolean variables in the $\ell$–th clause take the values opposite to the desired ones, i.e. iff the clause itself is false. The above expression can also be written in a way which is manifestly reminiscent of spin-glass models (and more precisely neural networks with an extended Hebbian rule [2]),

$$E[\Delta, S] = \frac{\alpha}{2^K} N + \sum_{R=1}^{K} (-1)^R \sum_{i_1 < i_2 < \ldots < i_R} J_{i_1, i_2, \ldots, i_R} S_{i_1} S_{i_2} \ldots S_{i_R} \quad , \tag{3}$$

where the couplings are defined by

$$J_{i_1, i_2, \ldots, i_R} = \frac{1}{2^K} \sum_{\ell=1}^{M} \Delta_{\ell, i_1} \Delta_{\ell, i_2} \ldots \Delta_{\ell, i_R} \quad . \tag{4}$$

In view of the above formulation and of the current knowledge on long-range spin–glasses, we may already expect qualitatively different behaviours for $K = 2$ (similar to Sherrington-Kirkpatrick model) and $K \geq 3$ (closer to the so-called p-spins or Potts models) [1]. We shall see in the following that analytical calculations support this intuitive feeling.

Finally, to ensure that the number of Boolean variables in any clause is exactly equal to $K$, we impose on $\Delta$ the following constraints

$$\sum_{i=1}^{N} \Delta_{\ell,i}^2 = K, \ \ \forall \ell = 1, ..., \ M \quad . \tag{5}$$

The ground state (GS) properties of the cost function (2) will reflect those of K–SAT ($E_{GS} = 0$) and MAX–K–SAT ($E_{GS} > 0$). In (2), one may interpret $K$ as the number of "neighbours" to which each spin is coupled inside a clause. To study the ground state properties fo the cost function (2), we follow the replica approach in the framework of diluted models which is indeed much more complicated than that of long–range fully connected disordered models. As we shall see below, replica theory must be formulated in a functional form involving not only interactions between pairs of replicas but all multi–replicas overlaps. To be more precise, we shall use below a new order parameter formulation, inspired from [13,14], which results much more convenient than usual overlaps.

To compute the ground state energy, we first introduce a fictitious temperature $1/\beta$ to regularize all mathematical expressions and send $\beta \to \infty$ at the end of the calculation. Note that the introduction of a finite temperature also greatly helps to understand the physical properties of the model. We proceed by computing the model "free–energy" density at inverse temperature $\beta$, averaged over the clauses distribution

$$F(\beta) = -\frac{1}{\beta N}\overline{\ln Z[\Delta]} \quad , \tag{6}$$

where $Z[\Delta]$ is the partition function

$$Z[\Delta] = \sum_{\{S_i\}} \exp\left(-\beta E[\Delta, S]\right) \quad . \tag{7}$$

As already mentioned, the energy (2) is self-averaging and can therefore be obtained from the above free–energy. The overline denotes the average over the random clauses matrices satisfying the constraint (5) and is performed using the replica trick $\overline{\ln Z} = \lim_{n\to 0} \frac{\overline{Z^n}-1}{n}$, starting from integer values of $n$. The typical properties of the ground state, i.e. the internal energy and the entropy, will then be recovered in the $\beta \to \infty$ limit.

Once averaged over the clauses choices, the $n^{th}$ integer moment of the partition function depends on the spins only through the multi-overlaps

$$Q^{a_1,a_2,\ldots,a_{2r}} = \frac{1}{N}\sum_{i=1}^{N} S_i^{a_1} S_i^{a_2} \ldots S_i^{a_{2r}} \quad , \tag{8}$$

involving an even number of replicas. To avoid the introduction of conjugated Lagrange parameters, we introduce along the lines of [13,14] a new generating function

$$c(\vec{\sigma}) = \frac{1}{2^n}\left(1 + \sum_{r=1}^{n/2} \sum_{a_1<\ldots<a_{2r}} Q^{a_1,a_2,\ldots,a_{2r}} \sigma^1\sigma^2\ldots\sigma^n\right) \quad , \tag{9}$$

where $\vec{\sigma} = (\sigma^1, \sigma^2, \ldots, \sigma^n)$ spans the space of all $2^n$ vectors with $n$ binary components $\sigma^a = \pm 1$. The use of this new order parameter lead to simpler algebraic calculations than the usual procedure involving the overlaps (8) and their Lagrange multipliers. Its physical interpretation is straightforward : $c(\vec{\sigma})$ equals the fraction of sites $i$ (among all possible $N$ Boolean variables) such that $S_i^a = \sigma^a$, $\forall a = 1, \ldots, n$. Therefore, all $c(\vec{\sigma})$'s range from zero and one and the global normalization condition implies that

$$\sum_{\vec{\sigma}} c(\vec{\sigma}) = 1 \qquad . \tag{10}$$

In addition, the vanishing condition on overlaps with an odd number of replica indices reads

$$c(\vec{\sigma}) = c(-\vec{\sigma}) \qquad , \qquad \forall\, \vec{\sigma} \qquad . \tag{11}$$

The averaged integer moments of the partition function are then given by the following formula

$$\overline{Z[\Delta]^n} = \int_0^1 \prod_{\vec{\sigma}} dc(\vec{\sigma})\; e^{N\ \mathcal{F}[\{c\},K,\alpha,\beta]} \qquad , \tag{12}$$

where the integration measure is restricted to $c(\vec{\sigma})$'s fulfilling constraints (10,11) and

$$\mathcal{F}[\{c\},K,\alpha,\beta] =$$
$$- \sum_{\vec{\sigma}} c(\vec{\sigma}) \ln c(\vec{\sigma}) + \alpha \ln \left[ \sum_{\vec{\sigma}_1,\ldots,\vec{\sigma}_K} c(\vec{\sigma}_1)\ldots c(\vec{\sigma}_K) \prod_{a=1}^{n} \left( 1 + (e^{-\beta} - 1) \prod_{\ell=1}^{K} \delta[\sigma_\ell^a ; 1] \right) \right] \qquad . \tag{13}$$

We may interpret the above free–energy functional as the free–energy of a $2^n$ interacting levels system. While the first term in $\mathcal{F}$ simply accounts for the statistical entropy, the second term represents the interactions between the levels at an effective “temperature” $1/\alpha$.

In the large $N, M$ limit (with fixed $\alpha = M/N$), the partition function (12) may be evaluated by taking the saddle-point over all order parameters $c$. Since the function $\mathcal{F}$ is invariant under permutation of replicas, a possible natural saddle-point can be sought within the so called replica symmetric (RS) Ansatz [11,9,13,14]

$$c(\sigma^1, \sigma^2, \ldots, \sigma^n) = C\left( \sum_{a=1}^{n} \delta[\sigma^a ; -1] \right) \qquad , \tag{14}$$

which preserves permutation invariance. Constraints (10,11) now read

$$\sum_{j=0}^{n} \binom{n}{j} C(j) = 1 \qquad \text{and} \qquad C(n-j) = C(j) \qquad (0 \le j \le n) \qquad . \tag{15}$$

We obtain $n+1$ saddle–point equations for all $C(j)$'s by differentiating equation (13) with respect to the order parameters. In the $n \to 0$ limit, we are therefore provided with an infinity of order parameters $C(j)$ for any real number $j$. To reach a simple final expression of the order parameters, we now adopt the functional formalism proposed in [11,12]. Let us call $P(x)$ the (even) probability distribution of the Boolean magnetizations $x = \langle S \rangle$, averaged over the disorder $\Delta$. We show in Appendix that

$$C(j) = \int_{-1}^{1} dx \; P(x) \; \left( \frac{1-x}{1+x} \right)^j \tag{16}$$

in the limit $n \to 0$. The advantage of the above formulation is that $P(x)$ has a clear significance, directly comparable to numerical simulations. We shall come back on this point in next Sections.

After some algebra, we find the self-consistent equation for the magnetizations distribution $P(x)$ taking into account saddle–point conditions for all $C(j)$'s,

$$P(x) = \frac{1}{1-x^2} \int_{-\infty}^{\infty} du \, \cos\left[ \frac{u}{2} \ln\left( \frac{1+x}{1-x} \right) \right] \times \\ \exp\left[ -\alpha K + \alpha K \int_{-1}^{1} \prod_{\ell=1}^{K-1} dx_\ell P(x_\ell) \cos\left( \frac{u}{2} \ln A_{(K-1)} \right) \right] \tag{17}$$

with

$$A_{(K-1)} \equiv A_{(K-1)}(\{x_\ell\}, \beta) = 1 + (e^{-\beta} - 1) \prod_{\ell=1}^{K-1} \left( \frac{1+x_\ell}{2} \right) \qquad . \tag{18}$$

The corresponding replica symmetric free–energy reads

$$-\beta F(\beta) = \ln 2 + \alpha(1-K) \int_{-1}^{1} \prod_{\ell=1}^{K} dx_\ell P(x_\ell) \ln A_{(K)} + \\ \frac{\alpha K}{2} \int_{-1}^{1} \prod_{\ell=1}^{K-1} dx_\ell P(x_\ell) \ln A_{(K-1)} - \frac{1}{2} \int_{-1}^{1} dx P(x) \ln(1-x^2) \qquad . \tag{19}$$

Note that in eq.(19) $A_{(K)}$ si given by a formula similar to (18), where the upper bound of the product is replaced by $K$. To end with, let us remark that equation (17) can in turn be transformed into an integro–differential equation

$$\frac{\partial P(x)}{\partial \alpha} = -KP(x) + K \int_{-1}^{1} \prod_{\ell=1}^{K-1} dx_\ell \left[ P(x_1) + \alpha(K-1) \frac{\partial P(x_1)}{\partial \alpha} \right] P(x_2) \dots P(x_{K-1}) \times$$
$$\frac{1}{2} \left[ \frac{\partial \eta(x)}{\partial x} P(\eta(x)) + \frac{\partial \eta(-x)}{\partial x} P(\eta(-x)) \right] \tag{20}$$

where $\eta(x) = [(x+1)A_{(K-1)} + x - 1]/[(1+x)A_{(K-1)} + 1 - x]$ and for which the boundary condition is given by the solution of (17) in $\alpha = 0$:

$$P(x)|_{\alpha=0} = \delta(x) \quad . \tag{21}$$

## IV. A TOY MODEL : THE $K = 1$ CASE

The $K = 1$ case can be solved either by a direct combinatorial method or within our statistical mechanics approach. Though this particular case does not present any critical behaviour, its study will turn out to be useful in understanding the $K > 1$ models in which we are interested. Moreover, the $K = 1$ toy model allows to check the correctness of the statistical mechanics results.

In this case, a sample of $M$ clauses is completely described by giving directly the numbers $t_i$ and $f_i$ of clauses imposing that a certain Boolean variable $S_i$ must be true or false respectively. Therefore the partition function corresponding to a given sample reads

$$Z[\{t, f\}] = \prod_{i=1}^{N} (e^{-\beta t_i} + e^{-\beta f_i}) \quad , \tag{22}$$

and the average over the disorder gives

$$\begin{aligned} \frac{1}{N} \overline{\ln Z[\{t, f\}]} &= \frac{1}{N} \sum_{\{t_i, f_i\}} \frac{M!}{\prod_{i=1}^{N} (t_i! f_i!)} \ln Z[\{t, f\}] \\ &= \ln 2 - \frac{\alpha\beta}{2} + \sum_{l=-\infty}^{\infty} e^{-\alpha} I_l(\alpha) \ \ln \left( \cosh \left( \frac{\beta l}{2} \right) \right) \quad , \end{aligned} \tag{23}$$

where $I_l$ denotes the $l^{th}$ modified Bessel function. The zero temperature limit gives the ground state energy

$$E_{GS}(\alpha) = \frac{\alpha}{2} [1 - e^{-\alpha} I_0(\alpha) - e^{-\alpha} I_1(\alpha)] \tag{24}$$

and the ground state entropy

$$S_{GS}(\alpha) = e^{-\alpha} I_0(\alpha)\ \ln 2 \quad . \tag{25}$$

One may notice that for any $\alpha > 0$, the ground–state energy is positive. Therefore, the clauses are never satisfiable all together and the overall function (1) is false with probability one. Nonetheless, the entropy is finite, implying an exponential degeneracy of the ground–state describing the minimum number $N \cdot E_{GS}(\alpha)$ of unsatisfiable clauses. Such a degeneracy is due to the presence of a finite fraction of variables $e^{-\alpha} I_0(\alpha)$ which are subject to equal opposite constraints $t_i = f_i$, and whose corresponding spins may be chosen up or down indifferently without changing the energy.

The above results are indeed recovered in our approach, showing that the RS Ansatz is exact for all $\beta$ and $\alpha$ when $K = 1$. Equation (17) can be explicitly solved at any temperature $1/\beta$ and the solution reads

$$P(x) = \sum_{\ell=-\infty}^{\infty} e^{-\alpha} I_\ell(\alpha)\ \delta\left(x - \tanh\left(\frac{\beta \ell}{2}\right)\right) \quad , \tag{26}$$

which, in the limit of physical interest $\beta \to \infty$, becomes

$$P(x) = e^{-\alpha} I_0(\alpha) \delta(x) + \frac{1}{2}(1 - e^{-\alpha} I_0(\alpha))\left(\delta(x-1) + \delta(x+1)\right) \quad . \tag{27}$$

The finite value of the ground state entropy may be ascribed to the existence of unfrozen spins whose fractional number is simply the weight of the $\delta$–function in $x = 0$. At the same time, it appears that the non zero value of the ground state energy is due to the presence of completely frozen spins of magnetizations $x = \pm 1$. This is an important feature of the problem which remains valid for any $K$, as we shall see in the following. In Fig. 1 we report the plots of the above energy and entropy at zero temperature.

## V. REPLICA SYMMETRIC SOLUTIONS FOR ALL $K$

A relevant general mechanism for the comprehension of the overcoming critical behaviour in K–SAT is the accumulation of Boolean magnetizations $\overline{\langle S \rangle} = \pm(1 - O(e^{-|z|\beta}))$, $(z = O(1))$, in the limit of zero temperature and for $\alpha \to \alpha_c$. The emergence of Dirac peaks in $x = \pm 1$ signals that a freezing process has just occurred and that a further increase of $\alpha$ beyond $\alpha_c$ would cause the appearance of unsatisfiable clauses. This scenario – which can be verified by inspection of eq. (26) for $K = 1$ – is true also for $K > 1$. In fact, by computing the fraction of violated clauses through

$$E = -\frac{1}{N}\frac{\partial}{\partial \beta}\overline{\ln Z[\Delta]} \quad , \tag{28}$$

at temperature $1/\beta$, one sees that the ground state energy depends only upon the magnetizations of order $\pm(1 - O(e^{-|z|\beta}))$, if any, and that such contributions can be described by the introduction of the new rescaled function

$$R(z) = \lim_{\beta \to \infty} \left[ P\left(\tanh\left(\frac{\beta z}{2}\right)\right) \frac{\partial}{\partial z} \tanh\left(\frac{\beta z}{2}\right)\right] \quad , \tag{29}$$

whose meaning will be clarified in Section IX. ¿From (17), $R(z)$ fulfills the saddle–point equation

$$R(z) = \int_{-\infty}^{\infty} \frac{du}{2\pi} \cos(uz) \exp\left[ -\frac{\alpha K}{2^{K-1}} + \alpha K \int_0^\infty \prod_{\ell=1}^{K-1} dz_\ell R(z_\ell) \cos(u \, \min(1, z_1, \ldots, z_{K-1}))\right] \quad . \tag{30}$$

The corresponding ground state energy reads, see (19) and (29 ),

$$\begin{aligned} E_{GS}(\alpha) = \alpha(1-K) \int_0^\infty \prod_{\ell=1}^{K} dz_\ell R(z_\ell) \min(1, z_1, \ldots, z_K) + \\ \frac{\alpha K}{2} \int_0^\infty \prod_{\ell=1}^{K-1} dz_\ell R(z_\ell) \min(1, z_1, \ldots, z_{K-1}) - \int_0^\infty dz R(z) z \quad . \end{aligned} \tag{31}$$

It is easy to see that the saddle–point equation (30) is in fact a self–consistent identity for $R(z)$ in the range $z \in [0, 1]$ only. Outside this interval, equation (30) is merely a definition of the functional order parameter $R$. This remark will be useful in the following.

To start with, $R(z) = \delta(z)$ is obviously a solution of (30) for all values of $\alpha$ and $K$, giving a zero ground state energy since no spins are frozen with magnetizations $\pm 1$. Let us assume that $R(z)$ includes another Dirac peak in $0 < z_0 \leq 1$. Then, inserting this distribution in the exponential term on the r.h.s. of (30), we find that $R(z)$ on the l.h.s. necessarily includes all Dirac peaks centered in $kz_0$, where $k = 0, \pm1, \pm2, \pm3, \ldots$. Next, we proceed iteratively by inserting again the whole series in the r.h.s. of (30). For large enough $k$, $kz_0$ is larger than one and the exponentiated term includes a $\cos u$ contribution, which causes the presence of Dirac distributions centered in all (positive and negative) integers. Therefore, as soon as $R(z)$ is different from $\delta(z)$, it contains an infinite set of Dirac functions peaked around all integer numbers. Clearly, the simplest self–consistent solution to (30) will be obtained for $z_0 = 1$ since the process described above closes after one iteration. This solution reads [5]

$$R(z) = \sum_{\ell=-\infty}^{\infty} e^{-\gamma_1} I_\ell(\gamma_1) \delta(z-\ell) \quad , \tag{32}$$

where $\gamma_1$ depends on $K$ and $\alpha$ and fulfills the implicit equation

$$\gamma_1 = \alpha K \left[ \frac{1 - e^{-\gamma_1} I_0(\gamma_1)}{2} \right]^{K-1} \quad . \tag{33}$$

The physical meaning of $\gamma_1$ may be understood by looking at the definition of the rescaled function order parameter (29). Turning back to the magnetization distribution, we indeed find in the zero temperature limit

$$P(x) = e^{-\gamma_1} I_0(\gamma_1) P_r(x) + \frac{1}{2}(1 - e^{-\gamma_1} I_0(\gamma_1)) \left( \delta(x-1) + \delta(x+1) \right) \quad , \tag{34}$$

where $P_r(x)$ is a regular (i.e. without Dirac peaks in $x = \pm1$) magnetization distribution normalized to unity. The above identity is a straightforward extension of the expression (27) (when $K = 1$, $\gamma_1 = \alpha$ from (33) and $P_r(x) = \delta(x)$) to any value of $K$. Inserting eq.(32) in (31) gives the value of the cost–energy

$$E_{GS}(\alpha) = \frac{\gamma_1}{2K} \left( 1 - e^{-\gamma_1} I_0(\gamma_1) - K e^{-\gamma_1} I_1(\gamma_1) \right) \quad . \tag{35}$$

It is therefore clear that, in the RS context, the SAT to UNSAT transition corresponds to the emergence of peaks centered in $x = \pm1$ with finite weights, that is to a transition from $\gamma_1 = 0$ to $\gamma_1 > 0$. This simplest solution centered on integer numbers, similar to previous findings [11,12,26], was presented in ref. [5].

In addition to (32), there exist other RS solutions to the saddle–point equations [27]. For instance, if we choose $z_0 = \frac{1}{2}$, the insertion process ends up after two iterations and generates Dirac peaks centered in all integer and half–integer numbers. More generally, for any integer $p \geq 1$, we may define the solution to (30)

$$R(z) = \sum_{l=-\infty}^{\infty} r_\ell \, \delta \left( z - \frac{\ell}{p} \right) \quad , \tag{36}$$

having exactly $p$ peaks in the interval $[0,1[$, whose centers are $z_\ell = \frac{\ell}{p}$, $\ell = 0, \ldots, p-1$. The coefficients $r_\ell$ of these distributions are self–consistently found through

$$r_\ell = \int_0^{2\pi} \frac{d\theta}{2\pi} \cos(\ell\theta) \exp\left( \sum_{j=1}^{p} \gamma_j (\cos(j\theta) - 1) \right) \tag{37}$$

for all $\ell = 0, \ldots, p-1$ where

$$\begin{aligned} \gamma_j &= \alpha K \left[ \left( \frac{1}{2} - \frac{r_0}{2} - \sum_{\ell=1}^{j-1} r_\ell \right)^{K-1} - \left( \frac{1}{2} - \frac{r_0}{2} - \sum_{\ell=1}^{j} r_\ell \right)^{K-1} \right] \quad , \forall j = 1, \ldots, p-1 \\ \gamma_p &= \alpha K \left( \frac{1}{2} - \frac{r_0}{2} - \sum_{\ell=1}^{p-1} r_\ell \right)^{K-1} \quad . \end{aligned} \tag{38}$$

The corresponding energy reads, from (36) and (31),

$$\begin{aligned} E_{GS} = &\frac{\alpha(1-K)}{p} \left[ \left( \frac{1-r_0}{2} \right)^K + \sum_{j=1}^{p-1} \left( \frac{1-r_0}{2} - \sum_{l=1}^{j} r_l \right)^K \right] + \\ &\frac{\alpha K}{2p} \left[ \left( \frac{1-r_0}{2} \right)^{K-1} + \sum_{j=1}^{p-1} \left( \frac{1-r_0}{2} - \sum_{l=1}^{j} r_l \right)^{K-1} \right] - \sum_{j=1}^{p} \frac{j}{p} \gamma_j \left( \frac{r_0}{2} + \frac{r_j}{2} + \sum_{l=1}^{j-1} r_l \right) \quad . \end{aligned} \tag{39}$$

Note that the last term of (39) includes the coefficient $r_p$, which may be computed using identity (37). It is easy to check that the first non trivial solution (32) corresponds to $p = 1$. Though there might be continuous solutions to (30), we believe they can be reasonably approximated by the large $p$ solutions we have presented here [27]. In the following sections, we shall therefore analyze which are the physical implications of the above solutions in the different cases of interest, $K = 2$, $K \geq 3$ and $K >> 1$.

## VI. THE $K = 2$ CASE

The case $K = 2$ is the first relevant instance of K–SAT. Graph theory has allowed [20] to show that for $\alpha = \alpha_c(2) = 1$ the problem undergoes a satisfiability transition which can be also viewed as a P/ NP–complete transition, from 2–SAT to MAX–2–SAT.

Let us first consider the simplest $p = 1$ RS solution [5]. Self–consistency equation (33) leads to the solution $\gamma_1 = 0$ for any $\alpha$. However, for $\alpha > 1$ one finds another solution $\gamma_1(\alpha) > 0$ which maximizes the free–energy (and $E_{GS}$) and therefore must be chosen (this is a well known peculiar aspect of the replicas formalism [1]). When approaching the threshold from upside, we indeed find

$$E_{GS}(\alpha|p=1) = \frac{4}{27}\,(\alpha-1)^3 + O\left((\alpha-1)^4\right) \simeq 0.1481\,(\alpha-1)^3 \qquad . \tag{40}$$

As expected, the $p = 1$ RS theory predicts $E_{GS} = 0$ for $\alpha \leq 1$ and $E_{GS} > 0$ when $\alpha > 1$, giving back the rigorous result $\alpha_c(2) = 1$ : for $\alpha > 1$ the fraction of violated clauses becomes finite and the corresponding CNF formulas turn out to be false with probability one. The transition taking place at $\alpha_c$ is of second order with respect to the order parameter $\gamma_1$ and is accompanied by the progressive appearance of two Dirac peaks for $P(x)$ in $x = \pm 1$ with equal amplitudes $(1 - e^{-\gamma_1} I_0(\gamma_1))/2$.

It is straightforward to verify that RS solutions with $p \geq 2$ are not present below $\alpha = 1$. However, above the threshold, one has to check whether their ground state energy are larger than the one of the $p = 1$ solution, that is if they can be relevant for MAX–K–SAT. For $p = 2$, resolution of equations (37) and (38) close to $\alpha_c(2)$ leads to (discarding the choice $r_1 = 0$ which amounts to the $p = 1$ solution)

$$\begin{aligned} r_0 &= 1 - \frac{8+2\sqrt{2}}{7}(\alpha-1) + O\left((\alpha-1)^2\right) \\ r_1 &= \frac{3-\sqrt{2}}{7}(\alpha-1) + O\left((\alpha-1)^2\right) \end{aligned} \tag{41}$$

for the coefficients of the Dirac peaks in $z = 0$ and $z = \frac{1}{2}$ respectively. Inserting these expansion into the energy (39), one finds

$$E_{GS}(\alpha|p=2) = \frac{9+4\sqrt{2}}{98}\ (\alpha-1)^3 + O\left((\alpha-1)^4\right) \simeq 0.1496\ (\alpha-1)^3 \qquad , \tag{42}$$

which is slightly larger than the $p=1$ result (40). Numerical calculations for higher values of $p \geq 3$ confirm that the energy increases very slowly with $p$. We have found that for large $p$'s the ground state energy is almost stationary, so that the $p=10$ solution can be considered as a very fair approximation of the optimal $p \to \infty$ RS solution. The coefficients $r_\ell$ of the distributions present in the order parameter $R(z)$ (36) are displayed Fig. 2 for different values of $\alpha$ and in the cases $p=1$, $p=5$ and $p=10$.

The ground state energy predicted by the $p=10$ RS solution is compared to numerical exhaustive simulations carried out for small sized systems on Fig 3. For $\alpha > \alpha_c = 1$, the theoretical estimate of $E_{GS}$ seems to sligthly deviate from the numerical findings, which signals the occurrence of a Replica Symmetry Breaking (RSB) transition at the threshold. This is in agreement with a stability calculation performed on the Viana-Bray model [30] around the critical point $\alpha = 1$ [14]. Note that the Viana-Bray energy is, up to the (irrelevant at zero temperature) random field in (3) equivalent to the 2-SAT cost function. We may therefore expect that the result derived in [14] apply to our case. If it were so, there would be an instability of the replica symmetric saddle–point at the threshold due to replicon–like fluctuations, breaking replica symmetry above $\alpha_c$. The situation would be reminiscent of the case of neural networks with continuous weights, where RS theory is able to localize the storage capacity but not to predict the minimal fraction of errors beyond the transition [2,29]. The $1/N$ extrapolation of the simulations results from finite systems to $N \to \infty$ is shown Fig. 4 for the particular choice $\alpha = 3$. Data seem in favor of RSB but one cannot exclude that $1/N^2$ effects could make coincide both numerics and theory. However, one should notice that for $\alpha \gg 1$, the exact asymptotic scaling of the ground state energy $E_{GS} \simeq \alpha/4$ [24] is compatible with the RS prediction.

¿From the above discussion, it is reasonable to conclude that RS theory is exact in the region $0 \leq \alpha \leq 1$. As already mentioned, the key quantity to study in this range is the typical number of solution to the problem, i.e. the typical ground state entropy $S_{GS}(\alpha)$ given by eq.(19) in the $\beta \to \infty$ limit. Notice that a simpler expression of the ground state entropy, more precisely of its derivative, may be obtained by differentiating (19) with respect to $\alpha$ and using the saddle-point equation (20). The result reads

$$\frac{\partial S_{GS}}{\partial \alpha}(\alpha) = \int_{-1}^{1} \prod_{\ell=1}^{K} dx_\ell \; P(x_\ell) \; \ln \left[ 1 - \prod_{\ell=1}^{K} \left( \frac{1+x_\ell}{2} \right) \right] \qquad , \tag{43}$$

and is valid for any $\alpha$ and $K$. Using the initial value $S_{GS}|_{\alpha=0} = \ln 2$ and the above equation (43), one can in principle compute the ground state entropy for any value of $\alpha$. However, due to the difficulty of finding a solution of the integral equation (17), it turns out to be convenient to develop a systematic expansion of the entropy in the parameter $\alpha$. We now briefly present the procedure to be employed for a generic value of $K$.

Inserting $P(x)|_{\alpha=0} = \delta(x)$ into formula (43), we obtain the slope of the entropy at the origin

$$\left.\frac{\partial S_{GS}}{\partial \alpha}\right|_{\alpha=0} = \ln \left( 1 - \frac{1}{2^K} \right) \qquad , \tag{44}$$

which coincides with the annealed result [3,21]. Then, we use eq.(20) to compute the first derivative of the magnetizations distribution in $\alpha = 0$,

$$\left.\frac{\partial P(x)}{\partial \alpha}\right|_{\alpha=0} = -\alpha K \delta(x) + \frac{\alpha K}{2} \, \delta \left( x + \frac{1}{2^K - 1} \right) + \frac{\alpha K}{2} \, \delta \left( x - \frac{1}{2^K - 1} \right) \qquad . \tag{45}$$

Now, we differentiate eq.(43) with respect to $\alpha$ and inject the above result, which is needed to obtain the second derivative of the ground state entropy at $\alpha = 0$,

$$\left.\frac{\partial^2 S_{GS}}{\partial \alpha^2}\right|_{\alpha=0} = -K^2 \ln \left( 1 - \frac{1}{2^K} \right) + \frac{K^2}{2} \ln \left( 1 - \frac{1}{2^K - 1} \right) + \frac{K^2}{2} \ln \left( 1 - \frac{2^{K-1} - 1}{2^{K-1}(2^K - 1)} \right) \qquad , \tag{46}$$

which is negative as required since the entropy is expected to be a concave function of $\alpha$. The whole procedure, consisting in successive differentiations of eqs.(20) and (43) can then be iterated to compute symbolically all the derivatives of $P(x)$ and $S_{GS}(\alpha)$ with respect to $\alpha$ in $\alpha = 0$.

In the $K = 2$ case, we have calculated the power expansion of $S_{GS}(\alpha)$ up to the seventh order in $\alpha$ (which shows an uncertainty less than one percent with respect to the sixth order Taylor expansion on the range $\alpha \in [0;1]$). The result reads

$$S_{GS}(\alpha) = \ln 2 - 0.28768207\,\alpha - 0.01242252\,\alpha^2 - 0.0048241588\,\alpha^3 - 0.0023958362\,\alpha^4 - 0.0013119155\,\alpha^5 - 0.00081617226\,\alpha^6 - 0.00053068034\,\alpha^7 - \dots \quad , \tag{47}$$

in which, for simplicity, we have reported only few significant digits of the coefficients. The latter are computed symbolically and have the form of a logarithm of rational number. At the transition we find $S_{GS}(\alpha_c) \simeq 0.38$ which is indeed very high as compared to $S_{GS}(0) = \ln 2$. A plot of the entropy versus $\alpha$ is shown Fig. 3. For completeness, we stress that the ground state entropy and the logarithm of the number of solutions, which coincide below $\alpha_c$, have different meanings (and values) above the threshold. In this region, the latter equals to $-\infty$ since all solutions have disappeared while the former quantity reflects the degeneracy of the lowest state (with strictly positive energy) and is continuous at the transition as shown by simulations.

Since, for $\alpha > \alpha_c$, there do not exist anymore sets of $S_i$'s such that the energy (2) remains nonzero, the vanishing of the exponentially large number of solutions that were present below the threshold is surprisingly abrupt. We then conclude that the transition itself is due to the appearance, with probability one, of contradictory logical loops in all the solutions and not to a progressive disappearance of the number of these solutions down to zero. This perfectly agrees with the graph–theoretical derivation of the critical $\alpha$ which is indeed based on a probabilistic calculation of appearance of contradictory cycles in oriented random graphs representing Boolean formulas.

## VII. THE $K \geq 3$ CASE

The $K = 3$ case is the first NP–complete instance of K–SAT. The resolution of the RS equations leads to a scenario different from the previous $K = 2$ case. We shall see below that RS theory does not allow to derive the value of the threshold $\alpha_c(3) \simeq 4.2$, which was estimated by means of finite–size scaling techniques [3]. This is due to the fact that the calculation of $\alpha_c(3)$ requires the introduction of Replica Symmetry Breaking (RSB), leading to very complicated equations we have not yet succeeded in solving. However, it is a remarkable fact that, in the relevant region for 3–SAT, i.e. for $\alpha$ ranging from zero up to $\alpha_c(3)$, the ground state entropy computed using RS theory seems to be exact.

Let us start with the $p = 1$ RS solution (32). Solving eq. (33) leads to the following scenario (see Fig. 5). For $\alpha < \alpha_m(3) \simeq 4.667$, there exists the solution $\gamma_1 = 0$ only. At $\alpha_m(3)$, a non zero solution $\gamma_1(\alpha) \neq 0$ discontinuously appears. The corresponding ground state energy is negative in the range $\alpha_m(3) \leq \alpha < \alpha_s(3) = 5.181$, meaning that the new solution is metastable and that $E_{GS} = 0$ up to $\alpha_s(3)$. For $\alpha > \alpha_s(3)$ the $\gamma_1(\alpha) \neq 0$ solution becomes thermodynamically stable [33].

From the above scheme one is tempted to conclude that $\alpha_s(3)$ corresponds to the desired threshold $\alpha_c(3)$. However, this prediction is wrong since the experimental value $\alpha_c(3) \simeq 4.2$ is lower than both $\alpha_m(3)$ and $\alpha_s(3)$. The failure of the above $p = 1$ RS prediction is also confirmed by the large $K$ limit. One finds $\alpha_m(K) \sim K2^K/16/\pi$ and $\alpha_s(K) \sim K2^K/4/\pi$ which are larger than the exact asymptotic value $\alpha_c(K) \sim 2^K \ln 2$. It is worth noticing that (similarly to the $K = 2$ case) though the scaling of $\alpha_c(K)$ for large $K$ is wrong within the $p = 1$ RS Ansatz, the asymptotic value for large $\alpha$ (and any $K$) of the ground state energy for MAX–K–SAT is correctly predicted : $E_{GS}(\alpha) \sim \alpha/2^K$ [24].

We now turn to improved RS solutions by looking at larger values of $p$. When $p = 2$, the previous transition scenario remains qualitatively unaltered, but the precise values of the spinodal and the threshold points are quantitatively modified. One finds, see Fig. 5, that $\alpha_m(3|p = 2) \simeq 4.45$ while $\alpha_s(3|p = 2) \simeq 4.82$. The ground state energy curve is similar to the $p = 1$ curve but is shifted to the left. Though still incorrect, the $p = 2$ prediction is thus closer to the real threshold value. For larger integers $p$, we have found that $\alpha_m(3|p)$ and $\alpha_s(3|p)$ still decrease but quickly converge to the values 4.428 and 4.60 respectively (we observed a power low convergence by considering values of $p$ up to 30, see Table 1). In Fig. 6, we have plotted the values of the coefficients $r_\ell$ ($\ell = 0, \ldots, p-1$) entering (36) for $p = 1$, $p = 5$ and $p = 10$. The departure of the coefficient curves for $p = 5$ from the $p = 10$ curves displaying $r_0, r_2, r_4, r_6$ and $r_8$ is clearly visible as soon as the remaining coefficients of the $p = 10$ solution, namely $r_1, r_3, r_5, r_7$ and $r_9$ which are implicitly set to zero in the $p = 5$ solution, acquire a non negligible value.

The first order jump of the order parameters $\gamma_j$'s has a precise meaning in terms of the fraction of Boolean variables completely determined at the transition. We have seen that, in 2–SAT, the fraction of Boolean variables whose values cannot fluctuate in the different ground states, that is the heights of the Dirac peaks of $P(x)$ in $x = \pm 1$, progressively increases from zero when $\alpha$ crosses its critical value. For larger $K \geq 3$, there abruptly appears a finite fraction of the variables which are entirely constrained by the clauses fulfillment condition at the threshold. We can compute this critical fraction $f$ using the RS theory. From eq.(36), we simply obtain $f = 1 - r_0$. The $p = 1$ solution therefore gives $f \simeq 0.656$. Increasing $p$, the fraction of fixed variables at the threshold converge to $f \simeq 0.94$, see Table 1. Such a value is quantitatively consistent with the expected typical entropy ($S_{GS} \simeq 0.03$ at $\alpha = 4.60$) which may be easily converted into an upper bound for the fraction of fixed variables by the relation $S_{GS} \leq (1-f)\ln 2$, leading to $f < 0.96$. Moreover, numerical investigations confirm that a quite large fraction of the Boolean variables have the same value (either always true or false) in all satisfying logical assignments at the threshold $\alpha_c \simeq 4.2$ [28].

Therefore, we may conclude from the above analysis that RS theory is unable to correctly predict the value of the transition threshold but provides us with a sensible qualitative pattern of the SAT/UNSAT transition. When crossing the latter, a first order replica symmetry breaking transition presumably takes place. The calculation of the threshold value would require the introduction of a replica symmetry broken Ansatz to replace (14). However, the issue of RSB in diluted models is largely an open one [15], due to the complex structure of the saddle–point equations involved, and we shall not attempt here at pursuing in this direction.

In the following, we shall rather show that RS theory still provide a consistent and very precise analysis of the behaviour of the random K–SAT problem below its threshold. This requires the inspection of the ground state entropy in the region where $R(z) = \delta(z)$. Using the method exposed in the previous Section, we have computed $S_{GS}$ to the $8^{th}$ order in $\alpha$ and found that

$$S_{GS}(\alpha) = \ln 2 - 0.13353139\,\alpha - 0.00093730474\,\alpha^2 - 0.00011458425\,\alpha^3 - 0.000016252451\,\alpha^4 - 2.4481877\,10^{-6}\,\alpha^5 - 3.9910735\,10^{-7}\,\alpha^6 - 6.5447303\,10^{-8}\,\alpha^7 - 1.167915\,10^{-8}\,\alpha^8 - \ldots \quad , \tag{48}$$

in which, again, we have reported only few sufficient digits of the (exactly known) coefficients. The entropy curve is displayed Fig. 7 in the range $0 \le \alpha \le \alpha_c(3)$. By computing the zero entropy points ($\alpha_{ze}$) given by the $\ell - th$ order entropy expansion, one finds a convergent succession of values toward $\alpha_{ze}(3) = 4.75$ (within one percent of precision), definitely outside the range of validity $0 \le \alpha \le \alpha_s(3|p \to \infty) \simeq 4.60$ of the expansion (48). Notice that $\alpha_{ze}|_{\ell=1}(3) = 5.1909$ corresponds to the annealed theory. A similar calculation for the cases $K = 4, 5, 6$ yields qualitatively similar results which show an even quicker convergence towards a zero entropy point such that $\alpha_c(K) < \alpha_{ze}(K)$ (see next Section for the analysis of the large $K$ limit where both values coincide).

Therefore, $S_{GS}$ is always positive below $\alpha_s(3|p \to \infty)$. In contradistinction with the $p = 1$ RS solution [5], the large $p$ RS solution cannot be ruled out by a simple inspection of their corresponding entropy. A more important consequence of the previous calculation of the entropy is that, at the threshold $\alpha_c$, the RS entropy is still nonzero. The crucial point is now to understand whether such value of the entropy is exact up to $\alpha_c$ or whether Replica Symmetry Breaking (RSB) effects have come into play. This issue may be clarified by resorting to exhaustive numerical simulation. As reported in [5], simulations in the range $N = 12, ..., 28$ lead to the conclusion that not only the entropy is indeed finite at the transition but also that our analytical solution appears exact up to $\alpha_c$. In particular the $1/N$ extrapolation of the entropy value at $\alpha = 4.17$ shows a remarkable agreement between the numerical trend and the RS prediction $S_{GS}(\alpha_c) \simeq 0.1$ (see the inset of the figure in [5]). RSB corrections to the RS theory seem thus to be absent below $\alpha_c$, which leads us to conjecture that the RSB transition could occur at $\alpha_c$ exactly. In this sense the situation would be partially similar to the binary network case [34] : the RS entropy would be exact up to $\alpha_c$ (though without vanishing) that would also coincide with the symmetry breaking point. To end with, let us mention that the existence of an exponential number of solutions just below the threshold has been demonstrated [28]. The rigorous lower bound of $S_{GS}$ is $S_{min} \simeq 0.014$ (for 3–SAT), which is compatible with our result.

## VIII. THE ASYMPTOTIC CASE OF LARGE $K$

In the large $K$ limit, the saddle point equations lead to a closed form for the probability distribution $P(x)$. In fact, in terms of the quantity

$$Q(A) = \int_{-1}^{1} \prod_{\ell=1}^{K-1} dx_\ell P(x_\ell) \delta\left(A - A_{(K-1)}\right) \quad , \tag{49}$$

the differential equation (20) reads

$$\frac{\partial P(x)}{\partial \alpha} = \\ -KP(x) + K \int_{-\infty}^{\infty} dA \left( Q(A) + \alpha \frac{\partial Q(A)}{\partial \alpha} \right) . \frac{1}{2} \left[ \frac{\partial \eta(x)}{\partial x} P(\eta(x)) + \frac{\partial \eta(-x)}{\partial x} P(\eta(-x)) \right] , \tag{50}$$

where $\eta(x)$ has been defined in Section III. For $K \gg 1$ , we may expand $Q(A)$ as

$$Q(A) \simeq \delta(A-1) + \frac{1}{2^{K-1}} \delta'(A-1) + \frac{1}{2} \left( \frac{1}{4} + \frac{1}{4} \int_{-1}^{1} dx P(x) x^2 \right)^{K-1} \delta''(A-1) + \dots \tag{51}$$

Under the changes of variables $G(y,\alpha) = (1 - \tanh^2 y) P(\tanh y)$ and

$$V(\alpha) = \alpha K \left( \frac{1}{4} + \frac{1}{4} \int_{-1}^{1} dx P(x) x^2 \right)^{K-1} \quad , \tag{52}$$

equations (50) and (51) simplify into the celebrated heat equation

$$\frac{\partial G(y,V)}{\partial V} = 2 \, \frac{\partial^2 G(y,V)}{\partial y^2} \quad . \tag{53}$$

whose normalized solution is $G(y,V) = \exp(-y^2/2V)/\sqrt{2\pi V}$. Turning back to $P(x)$, we find

$$P(x) \simeq \frac{1}{\sqrt{2\pi V(\alpha)}(1-x^2)} \exp\left(-\frac{1}{8V(\alpha)} \ln^2\left(\frac{1+x}{1-x}\right)\right) \quad , \qquad (K \gg 1) \quad , \tag{54}$$

where $V(\alpha)$ is given by the self–consistency equation (52). The latter may be easily estimated for large $K$ : $V(\alpha) \simeq \alpha K/4^{K-1}$. Therefore, when $\alpha < \alpha_c(K) \simeq 2^K \ln 2$, $V(\alpha)$ is vanishingly small, that is $P(x) \to \delta(x)$, proving that the replicas become uncoupled in the large $K$ limit [3]. In addition, it can be checked that the zero entropy point $\alpha_{ze}(K)$ reaches the threshold $\alpha_c(K)$ from above. Another way of looking at the entropy is provided by equation (46) : it is a simple check the fact that $\alpha_c(K)^2 \frac{\partial^2 S_{GS}}{\partial \alpha^2}|_{\alpha=0} \to 0$ for large $K$. We may then conclude that the annealed approximation becomes exact when $K \gg 1$. As said above, $K$ may be understood as the connectivity of our model and, in the asymptotic regime $K \gg 1$, RS theory includes only Gaussian interactions as in long–range spin–glasses models [34]. In Fig. 8 we report some instances of the probability distribution, calculated for different values of $K$ and $\alpha$. Notice that since the critical point coincides, in this large $K$ limit, with the zero entropy point (which is far below the point where the RS energy becomes positive - see previous Section), the probability distribution of the Boolean magnetization is far from being concentrated in $\pm 1$.

## IX. ALTERNATIVE DERIVATION OF THE SELF-CONSISTENCY EQUATION FOR R(Z)

In this Section, we discuss an alternative heuristic derivation of the self-consistency equation (30) for $R(z)$ without resorting to replicas. As a result of this approach, we shall unveil the physical meaning of the $R(z)$ functional order parameter and interpret the replica symmetry assumption in probabilistic terms. The method we adopt is known as the cavity approach [1,26] and here we need to transpose it to the zero temperature case. For the sake of simplicity, we shall focus on the 2–SAT case, extensions to higher instances of $K$–SAT being straightforward.

To each Boolean variable $x_i$ and for a given logical formula, we associate a quantity $z_i$ defined as follows. We call $z_i$ the difference between the number of unsatisfied clauses $\mathcal{L}$ when $x_i = 0$ (false) and when $x_i = 1$ (true), averaged over the set of all optimal (for K–SAT or MAX–K–SAT) Boolean assignments, that is ground state configurations.

$$z_i = \mathcal{L}(x_i = 0) - \mathcal{L}(x_i = 1) \qquad . \tag{55}$$

Next, we consider the set of all $z_i$'s and define $T(z)$ as their probability distribution after having averaged over all possible logical formulas. The calculation of $T(z)$ proceeds according to the following four steps.

Let us consider a given Boolean variable, say $x_1$. **(I)** For uncorrelated random CNF expressions, the probability that neither $x_1$ nor $\bar{x}_1$ appear in the logical formula is simply $(1-\frac{2}{N})^M \simeq e^{-2\alpha}$. In such case, $x_1$ can be indifferently chosen either true or false, and $z_1 = 0$. Therefore, we obtain a first contribution

$$T_0(z_1) = e^{-2\alpha}\delta(z_1) \tag{56}$$

to $T(z_1)$. **(II)** With probability $2\alpha\, e^{-2\alpha}$, $x_1$ will belong to a single clause, e.g. $x_1 \vee \bar{x}_2$. The latter is unsatisfied if and only if $x_1$ is false and $x_2$ is true. Therefore, $z_1 = 0$ if $x_2$ is allowed to be false (the clause is satisfied independently on $x_1$), i.e. if $z_2 \leq 0$. In order to see what happens in the average case, let us consider the case where $x_2$ is true in the majority of optimal Boolean assignments. At first sight, $z_2$ would appear as a strictly positive integer since it coincides with a difference of integer numbers [11,12], leading to $z_1 = 1$. However, as we consider averaged differences, the $z_j$ may well be rational numbers [27]. Such a counter intuitive behavior can be easily understood with the following simple argument. If $x_2$ appears (in average) in less than one clause, that is if $0 < z_2 < 1$, it cannot be present in another clause and we must have $z_1 = z_2$. Conversely, if $z_2 > 1$, $x_2$ is more frozen than $x_1$ and $z_1$ saturates its upper bound equal to one. Notice that this result may be made rigorous by working at finite temperature [26]. To complete the probabilistic analysis of this second contribution $T_1(z_1)$ to $T(z_1)$, we have to take into account the three other possible clauses involving $x_1$ and $x_2$, and collect the corresponding contributions. We find

$$T_1(z_1) = 2\alpha\, e^{-2\alpha} \int_0^\infty dz_2\; T(z_2)\; \left\{ \frac{1}{2}\delta(z_1 - \; \min(1,z_2)) + \frac{1}{2}\delta(z_1 + \min(1,z_2)) + \delta(z_1) \right\} \; . \tag{57}$$

**(III)** By iterating the above reasoning, we consider a logical formula such that $x_1$ belongs exactly to $j$ clauses. The probability of such an event obeys the Poisson law $(2\alpha)^j e^{-\alpha}/j!$. Almost surely, the variables $x_2, x_3, \ldots, x_{j+1}$ appearing in these $j$ clauses are different from each other. Moreover, in the large $N$ limit, any pair of variables $x_m$ and $x_n$ $(2 \leq m,n \leq j+1)$ are always at a large "distance" from one another, where the relative distance is defined as the minimal number of logical links (clauses) joining $x_m$ to $x_n$, see ref. [26]. As a consequence, the joint probability distribution of $z_2, z_3, \ldots, z_{j+1}$ factorizes and due to the statistical independence of the choices of the clauses, we have

$$T_j(z_1) = \frac{(2\alpha)^j}{j!}\; e^{-2\alpha} \int_0^\infty \prod_{\ell=2}^{j+1} dz_\ell T(z_\ell) \sum_{m=0}^{j} \sum_{a_1<\ldots<a_m} \frac{1}{2^m} \sum_{\sigma_1,\ldots,\sigma_m=\pm 1} \delta\left( z_1 - \sum_{\ell=1}^{m} \sigma_\ell \, \min(1, z_{a_\ell}) \right) \; , \tag{58}$$

where the $a_\ell$'s run between 2 and $j+1$. **(IV)** Summing the previous expressions for all values of $j$, we recover eq.(30) with $R(z) = T(z)$.

Of course, the self–consistency equation for $R(z)$ is correct provided that replica symmetry is valid, while $T(z)$ is defined independently from any replica calculation. Therefore the equality between the two quantities cannot hold in general and is due to the assumption on the absence of correlations between different $z_\ell$ we have made above [1,26]. This is the probabilistic meaning of replica symmetry.

## X. CONCLUSION AND PERSPECTIVES

In this paper, we have presented the replica symmetric theory of the random K–SAT problem. We have shown that the natural quantity emerging from the analytical study is the distribution of the average values of the Boolean variables, indicating to what extent the latters are determined by the constraints imposed by the clauses. The knowledge of this probability distribution requires the resolution of a functional saddle–point equation, for which we have presented an iterative sequence of improved solutions. The most surprising result we have derived is the fact that the entropy is finite just below the transition, i.e. that the latter is characterized by an abrupt disappearance of all exponentially numerous solutions due to the emergence of contradictory loops.

Some numerical simulations we have performed for $K = 2$ as well as in the $K = 3$ case are in remarkable quantitative agreement with our RS calculations of the entropy jump at the threshold [5]. Both the known results on the stability of 2–SAT like models and the numerical simulations, hint at the correctness of the RS theory up to the critical ratio of clauses per Boolean variable.

Would it be so the physical picture of the space of solutions would not necessarily be simple. Replica symmetry can indeed hide a non trivial structure of the solutions, as has been shown for long range spin-glasses [35] models and in the (closer to K–SAT) case of neural networks [36]. This issue is probably of crucial importance to understand the performances of local search algorithms.

As for the values of the critical thresholds themselves, RS gives the correct prediction $\alpha_c = 1$ for $K = 2$ but fails in estimating the critical $\alpha_c$ for $K \geq 3$. The study of the (hard) instances $K \geq 3$ of the K–SAT problem requires to break replica symmetry. As a consequence, their direct study will not be easy and will require non trivial analytical efforts.

Another route which one can follow to reach a better understanding of the $K > 2$ case consists in starting from the relatively well understood 2–SAT case and modifying it to get closer to the 3–SAT problem. Such a perturbative approach can be implemented by considering a mixed model, which one may refer to as $(2 + \epsilon)$–SAT model ($\epsilon \in [0, 1]$), composed of $(1-\epsilon)M$ clauses of length three and $\epsilon M$ clauses of length two (thus interpolating smoothly between the Polynomial 2–SAT and the NP–complete 3–SAT models). Analytical investigations suggest that the threshold can be computed exactly up to $\epsilon = \epsilon_0 = 0.413$.

For $\epsilon \leq \epsilon_0$, one finds a continuous SAT/UNSAT transition at $\alpha_c(\epsilon) = 1/(1-\epsilon)$. The model shares the same physical features as the random 2–SAT model. For $\epsilon > \epsilon_0$ , the SAT/UNSAT transition becomes a discontinuous (with respect to the order parameters) RSB transition similarly to the 3–SAT model. Preliminary numerical results suggest that the above model can be of interest for exploring the connection between the nature of the RS to RSB phase transition and the onset of exponential regimes in search algorithms running on samples generated near criticality [31].

**Acknowledgments** : We thank O. Dubois, S. Kirkpatrick and B. Selman for useful discussions.

## APPENDIX A: RELATIONSHIPS BETWEEN ORDER PARAMETERS

Identity (9) implicitly implies that one can make a change of variables from overlaps $Q$ to the generating function $c$. Let us call $\mathcal{M}$ the linear operator

$$\mathcal{M}(\{a_1, a_2, \ldots, a_{2p}\}; \vec{\sigma}) = \sigma^{a_1}\sigma^{a_2}\ldots\sigma^{a_n} \qquad . \tag{A1}$$

For simplicity, we set $Q^{\{\emptyset\}} = 1$ and all overlaps with an odd number of replicas are null. The dimension of $\mathcal{M}$ is therefore equal to $2^n$. To any sequence $\{a_1, a_2, \ldots, a_n\}$, we associate a $n$-component vector $\vec{\tau}$ such that $\tau^b = -1$ if $b$ belongs to the sequence and $\tau^b = 1$ otherwise. From definition (A1), we obtain

$$\mathcal{M}(\vec{\tau}; \vec{\sigma}) = \prod_{a=1}^{n} \frac{1}{2}(1 + \sigma^a + \tau^a - \sigma^a\tau^a) \qquad . \tag{A2}$$

As a consequence, $\mathcal{M}$ equals the $n^{th}$ power (for tensor product) of a two by two matrix. The Jacobian of the change of variables is found to be

$$|\mathcal{M}| = (-2)^{n2^{n-1}} \tag{A3}$$

and is different from zero. We may invert $\mathcal{M}$ and find

$$c(\vec{\sigma}) = \frac{1}{2^n}\left(1 + \sum_{p=1}^{n/2} \sum_{a_1<a_2<\ldots<a_{2p}} Q^{a_1,a_2,\ldots,a_{2p}}\ \sigma^{a_1}\sigma^{a_2}\ldots\sigma^{a_{2p}}\right) \qquad . \tag{A4}$$

Let us now turn to the replica symmetric Ansatz structure. From definition (14) and identity (A4), we obtain

$$C(j) = \frac{1}{2^n} \left( 1 + \sum_{p=1}^{n/2} Q_p \sum_{a_1 < a_2 < \ldots < a_{2p}} \sigma^{a_1} \sigma^{a_2} \ldots \sigma^{a_{2p}} \right) \quad , \tag{A5}$$

where $2j = n - \sum_{a=1}^{n} \sigma^a$ and the replica symmetric overlaps $Q_p$ are calculated from the magnetizations distribution [11,12]

$$Q_p = \int_{-1}^{1} dx \; P(x) \; x^{2p} \quad . \tag{A6}$$

To establish the relationship between the $C(j)$'s and the distribution $P(x)$, we have to expand the sum over replicas taking place in (A5) onto the powers of the $\vec{\sigma}$ magnetization

$$\sum_{a_1 < a_2 < \ldots < a_{2p}} \sigma^{a_1} \sigma^{a_2} \ldots \sigma^{a_{2p}} = \sum_{r=0}^{p} H_{p,r}^{(n)} \left( \sum_{a=1}^{n} \sigma^a \right)^{2r} \quad . \tag{A7}$$

The matrix $H^{(n)}$ can be computed by first finding the generating function of $[H^{(n)}]^{-1}$ and then inverting the latter. We finally find

$$H_{p,r}^{(n)} = \frac{1}{(2p)!(2r)!} \frac{\partial^{2p}}{\partial y^{2p}} \left( \sqrt{1-y^2} \right)^n (\text{Arctanh } y)^{2r} \Big|_{y=0} \quad . \tag{A8}$$

Using the above expression and inserting eq.(A6) into (A5), one recovers identity (16) in the limit $n \to 0$.

## REFERENCES

* Email: monasson@physique.ens.fr ; preprint LPTENS 96/35

LPTENS is a unité propre du CNRS, associée à l'Ecole Normale Supérieure et à l'Université de Paris–Sud.

† Email: zecchina@to.infn.it; preprint POLFIS-TH 96/15

also at INFN sez. Torino.

| $p$ | $\alpha_s^{RS}$ | $f$ |
|---|---|---|
| 1 | 5.1812 | 0.6561 |
| 3 | 4.7271 | 0.7889 |
| 6 | 4.6451 | 0.8497 |
| 9 | 4.6240 | 0.8765 |
| 12 | 4.6153 | 0.8920 |
| 15 | 4.6107 | 0.9022 |
| 18 | 4.6080 | 0.9095 |
| 21 | 4.6063 | 0.9150 |
| 24 | 4.6051 | 0.9193 |
| 27 | 4.6042 | 0.9227 |
| 30 | 4.6036 | 0.9256 |

TABLE I. $p$ dependence of the RS critical ratio $\alpha_s^{RS}$ and of the fraction $f$ of fixed variables.

# FIGURES

FIG. 1. Ground state cost (bold line), or fraction of violated clauses, and entropy (thin line) versus $\alpha = M/N$ for $K = 1$.

FIG. 2. Order parameters $r_i$ ($i = 0, ..., p-1$) corresponding to the different RS solutions $p = 1$ (dashed line), $p = 5$ (dashed–dotted lines) and $p = 10$ (continuous lines), for $K = 2$ and $\alpha = M/N \in [1, 3]$. The upper curve within each group represent $r_0$ whereas the overlapping ones in the lower part of the figure represent $r_i$ for $i = 1..., p-1$ ($p = 5, 10$).

FIG. 3. RS ground state entropy (decreasing curve, left scale) and RS ground state cost (increasing overlapping curves computed for $p = 1, ..., 10$, right scale) versus $\alpha = M/N$ for $K = 2$. At $\alpha = \alpha_c = 1$ the ground state cost becomes positive, signaling a second order SAT/UNSAT transition (at the same point the RS solution becomes unstable). The value of the entropy at the critical ratio is 0.38. The dashed lines interpolate the numerical data of exhaustive simulations on systems of size $N = 16, 20, 24$ and averaged over $15000, 7500, 2500$ samples respectively. Errors bars are within 10% for the entropy and even smaller for the energy and thus not reported explicitly.

FIG. 4. $1/N$ extrapolation of the minimal fraction of violated clauses (i.e. ground state cost) for $\alpha = 3$ and $N = 18, 20, 22, 24, 26$ averaged over $20000, 15000, 10000, 7500$ and $5000$ samples respectively. The extrapolated value appears to be different from the value 0.14472 toward which the RS solutions with increasing $p$ rapidly converge. This is in agreement with the expected instability of the RS solutions for $\alpha > \alpha_c$.

FIG. 5. RS ground state energy for $K = 3$ (continuous lines) computed for $p = 1, ..., 10$ (lines corresponding to larger values of $p$ would not be distinguishable) and compared with the results of numerical simulations on systems of size $N = 16, 20, 24$ and averaged over $15000, 7500, 2500$ samples respectively (error bars are of the order of the size of the dots). The RS ground state energy becomes positive (for $p >> 1$) at $\alpha_s \simeq 4.60$ whereas the value at which the unstable solution appears is $\alpha_m \simeq 4.428$. Both values are grater than the numerical estimate of the critical ratio (4.2). Scope of the dashed line is to help the eye in following the expected, yet unkown, RSB behaviour of the ground state energy.

FIG. 6. Order parameters $r_i$ $(i = 0, ..., p-1)$ corresponding to the different RS solutions $p = 1$ (dashed line), $p = 5$ (dashed–dotted lines) and $p = 10$ (continuous lines), for $K = 3$ versus $\alpha = M/N$. Within each group of $p = 1, 5, 10$ curves, the upper one represent $r_0$ whereas the others represent $r_i$ $(i = 1..., p-1)$, in top-down order.

FIG. 7. RS entropy (continuous line) for $K = 3$ versus $\alpha = M/N$ compared with the results of exhaustive numerical simulations for $N = 16, 20, 24$ and averaged over $15000, 7500, 2500$ samples respectively (see also ref.[4]). Errors bars are within 10% and not reported explicitly.

FIG. 8. Probability distributions $P(x)$ as functions of the magnetization $x$, calculated for $\alpha = 2^K \ln 2$ (critical threshold in the $K >> 1$ limit) and for $K = 10, 12, 14, 16, 18$.